# Adaptive LVRT Settings Adjustment for Enhancing Voltage Security of Renewable-Rich Electric Grids

Chen Wang
Chetan Mishra
Dominion Energy
Richmond, VA 23220, USA

Reetam Sen Biswas
Anamitra Pal
School of ECEE
Arizona State University
Tempe, AZ 85281, USA

Virgilio A. Centeno
Bradley Department of Electrical
and Computer Engineering
Virginia Tech, Blacksburg,
VA 24060, USA

*Abstract*—**Inverter based renewable generation (RG), especially at the distribution level, is supposed to trip offline during an islanding situation. However, islanding detection is done by comparing the voltage and frequency measurements at the point of common coupling (PCC), with limits defined in the form of ride-through curves. Current practice is to use the same limit throughout the year independent of the operating conditions. This could result in the tripping of RG at times when the system is already weak, thereby posing a threat to voltage security by heavily limiting the load margin (LM). Conversely, heavily relaxing these limits would result in scenarios where the generation does not go offline even during an islanding situation. The proposed methodology focuses on optimizing low-voltage ride-through (LVRT) settings at selective RGs as a preventive control for maintaining a desired steady-state voltage stability margin while not sacrificing dependability during islanding. The proposed process is a multi-stage approach, in which at each stage, a subset of estimated poor-quality solutions is screened out based on various sensitivities. A full continuation power flow (CPFLOW) is only run at the beginning and in the last stage on a handful of remaining candidate solutions, thereby cutting down heavily on the computation time. The effectiveness of the approach is demonstrated on the IEEE 9-bus system.**

*Index Terms*--Adaptive control, Continuation power flow, Low-voltage ride-through, Renewable energy, Voltage security

## I. Introduction

Increasing penetration of RG resources has introduced major complexities [1] to bulk electric system operation. In the past, when the overall penetration was low, RGs were made to trip offline as quickly as possible during system disturbances. This backfired as the amount of RGs became significant as the utilities could no longer afford to lose them [2]. Eventually, time-dependent curves defining the operating limits in terms of frequency and voltage referred to as the ride through curves were introduced, letting RGs stay online through the disturbance as long as their limits were met.

Events that most negatively impact both steady-state and dynamical system stability are often characterized by low voltage situations and therefore, the LVRT limit will be the focus of this paper. LVRT, as the name suggests, defines the lower limit on the voltage seen at the PCCs of RGs [3]. These curves are meant to ensure a perfect balance between tripping offline during islanding scenarios [4] and not tripping offline otherwise. However, in reality, it is extremely hard to differentiate islanding from non-islanding through local measurements. The misidentification of an islanding situation can potentially result in a transmission system level event that triggers a loss of large amounts of RGs. The effect of this generation loss on transient stability has been previously studied[5]. There is also a major impact on voltage security. Load margin (LM), a metric for voltage security, refers to the maximum additional load the system is able to supply such that an equilibrium point exists [6]. The situation could be especially severe when a large percentage of the load is being supplied by such prone-to-trip RGs.

In order to quantify the voltage security, there have been various approaches for LM estimation. The CPFLOW [7] is the most accurate and reliable method. It involves tracing the surface in state-space on which the load flow solutions evolve as load increases and thus is computationally demanding. The look-ahead approach utilized in [8], is reasonably fast and accurate and operates by approximating the power-voltage (PV) curve at one or more critical buses by fitting a parabola on a few load flow solutions. Lately, holomorphic embedding techniques [9] have become popular which provide a closed-form expression for load flow solution as a function of loading. One of the challenges with these fast approaches is the structural change in the load flow equations mainly due to generators hitting their Q-limits as load increases. In this situation, the solution undergoes a continuous transition to a new solution curve. The system studied in this paper has loss of generation due to LVRT violation, which is also a structural change to the equations. However, this change is computationally more challenging as the solution *jumps* between two non-intersecting solution curves which can be far apart in state space. Based on the authors' knowledge, LVRT based loss of generation is yet to be considered in voltage security assessment (mainly LM estimation) studies.

Various methods for enhancing the LM such as generation re-dispatch, load control, and shunt capacitance placement have been proposed in the past [8], [10]. However, situations can arise when they are not very effective in avoiding loss of critical RGs [1]. In such cases, having a methodology to adapt the

LVRT settings along with changing system conditions could prove to be extremely effective. The authors acknowledge the criticality of such a decision and therefore the aim would be to achieve it through minimal adjustment of LVRT settings and doing it only when absolutely necessary. Even if not deployed online for real-time control, having such a tool would give reasonable insights into 1) areas of the system requiring dynamic VAR resources to avoid tripping of critical RGs; and 2) regional/seasonal candidate LVRT settings.

The existing researches focusing on adaptive LVRT control are mostly concentrating on the control scheme designed to enhance the RGs' LVRT capability [11]. Others do control strategy design for specific RG(s) to provide better reactive support [12] or analyze impacts of RGs' LVRT capability on system protection [13]. For the LVRT criterion, researchers in [14] attempted to summarize proper LVRT settings from real-world practices. However, [11]-[14] treat the LVRT settings as an existing standard; they do not attempt to adjust the criterion for more secure and efficient system operation.

In this paper, a comprehensive method is proposed to adaptively adjust the LVRT limits of all critical RGs to improve RG utilization and ensure voltage security using a scalable 5-step approach. The main outcomes of this paper are:
1. A novel voltage prediction method for the identification of critical RGs to heavily cut down the number of candidate solutions.
2. Methodology for screening out ineffective LVRT adjustment solutions with regards to LM enhancement.
3. A method to estimate required LVRT adjustments.

A modified IEEE 9-bus system is used to demonstrate the effectiveness and efficiency of the proposed method.

## II. IMPACTS OF RG LVRT VIOLATIONS ON LOAD MARGIN

### A. Voltage Security and Load Margin

The standard power flow model is shown in (1),
$$f(x, \lambda) = F(x) + b\lambda = 0 \qquad (1)$$
where $\lambda \in R$ is the total load scaling factor; $b \in R^{N_L}$ is the direction of load growth; $N_L$ is the number of loads; and $x = [V \; \theta]^T$ is the vector of the unknown system states including voltage magnitudes $V$ at all PQ buses and phase angles $\theta$ at all the buses.

The equilibrium points of the system are the states that satisfy (1) for a given $\lambda$. On continuously varying $\lambda$, the equilibrium point(s) trace one-dimensional solution curves in the state-space. Of particular interest is the curve passing through the operating point of the system, i.e. the equilibrium point. However, this curve does not exist for all values of $\lambda$. The maximum value of $\lambda$ through which the curve does pass gives the LM, which is a widely used metric for quantifying voltage security [8]. No solution exists if $\lambda$ is increased beyond the LM; this is referred to as a saddle node bifurcation (SNB). This point on the curve corresponding to $\lambda = $ LM is characterized by $\frac{\partial f}{\partial x}$ becoming singular. The graph of voltages vs. $\lambda$ along this solution curve is called the P-V curve.

### B. P-V Curve Under LVRT tripping

A typical LVRT curve (can be seen in [3]) represents a time-dependent lower limit on voltage magnitude at the PCC of an RG. The violation of this limit results in tripping the RG offline. Note that the standards governing these limits do provide flexibility in their settings. Therefore, the system planners are free to propose closed-form values for their LVRT settings. However, an extra-relaxed LVRT curve will result in local RG not coming offline when needed and therefore there is a limit to which these curves can truly be adjusted. But this limit changes with operating conditions and thus is hard to know. It is important to clarify here that since we are looking at the steady-state voltage stability, we will be focusing on adjusting the final value of the curve.

We know that the connection status of an RG is a discrete variable that is prone to change due to violation of its LVRT limit. It is to be noted that there exists another set of discrete variables corresponding to generators hitting their reactive power output upper or lower limits, which is considered in the analysis but omitted from presentation here. The overall system can be modeled as,
$$f(x, \lambda, Z_{RG}) = F(x, Z_{RG}) + b\lambda \qquad (2)$$
where $Z_{RG} \in N^{N_{RG} \times 1}$ represents the connection statuses (0–offline/1–online) of RGs; $N_{RG}$ is the number of RGs. The solution curves for such a system have discontinuities due to $Z_{RG}$ being a discrete variable. This can be seen from a typical P-V curve for a single RG system as shown in Figure. 1. As voltage magnitude at the RG bus decreases with increasing $\lambda$, the LVRT limit (red dash-dot line) is hit, resulting in tripping of the RG. This makes the voltage jump from the original P-V curve to the curve corresponding to a system with no RG on that bus. Usually, with the loss of more RGs, the successive curves have smaller LMs as the loss of RGs causes an equivalent increase in system demand. Since $Z_{RG}$ is dependent on the state values and LVRT limits, the $\lambda$ at the jump is determined by local voltage support and LVRT settings.

Another interesting phenomenon plaguing renewable-rich systems is loss of equilibrium immediately following a jump. This happens when the solution does not exist for the succeeding system for the value of $\lambda$ when the jump happens. Under such circumstances, the SNB of the succeeding solution curve cannot be reached. For example, for the system in Figure. 1, if the LVRT setting was reduced to 0.75 p.u., the jump will happen at $\lambda > 1.2$ resulting in voltage collapse.

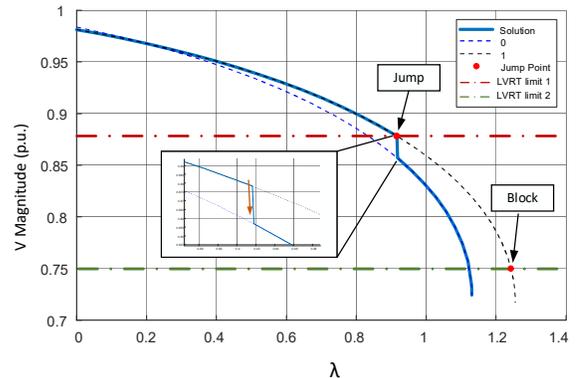

Figure. 1. Renewables tripping by violation of LVRT limit

## III. PROBLEM FORMULATION

The goal of this research is to achieve desired LM through adjustment of LVRT limits. Reaching $\lambda_{\text{limit}}$ means the existence of a solution of (2) at $\lambda = \lambda_{\text{limit}}$. This depends on the value of $Z_{RG}$ as $\lambda_{\text{limit}}$ is being approached, i.e. which curve to jump to. The only way to realize it is by ensuring the elements in $Z_{RG}$ corresponding to a certain set of RGs are equal to '1' at $\lambda = \lambda_{\text{limit}}$. This transforms the problem into finding that set of RGs which will be blocked from tripping. The new LVRT settings at those RGs will then be their voltage values at $\lambda = \lambda_{\text{limit}}$. Therefore, the proposed adaptive RG LVRT adjustment problem can be formulated as an optimization problem,

$$\underset{\Delta c_{LVRT}}{\operatorname{argmin}} \sum \Delta c_{LVRT}$$
$$s.t. \quad \Delta c_{LVRT}^{(i)} \geq 0, \ \forall i \in \{1, 2, \dots, N_{RG}\} \quad (a) \quad (3)$$
$$s.t. \quad \exists x \quad f(x, \lambda_{\text{limit}}, Z_{RG}(x, c_{LVRT})) = 0 \quad (b)$$

where $\Delta c_{LVRT} = c_{LVRT}^{original} - c_{LVRT}^{new} \in R^{N_{RG} \times 1}$ is the vector of LVRT adjustments for all the RGs in the system; $c_{LVRT}^{original} \in R^{N_{RG} \times 1}$ is the vector of original LVRT settings before any adjustment; $\lambda_{\text{limit}} \in R$ is the desired lower limit on LM. Since the connection status $Z_{RG}$ depends on the voltage magnitude and the LVRT setting, it can be written as:

$$Z_{RG}^{(i)} = \begin{cases} 0, & V_j < c_{LVRT}^{(i)} \\ 1, & \text{otherwise} \end{cases} \quad (4)$$

where the $i^{th}$ RG is installed on the $j^{th}$ bus; $V_j$ is the voltage magnitude of bus $j$; and $c_{LVRT}^{(i)}$ is the LVRT setting of RG $i$. The objective is to minimally adjust the LVRT settings of some RGs to block their unnecessary tripping when witnessing low voltages. Doing so will ensure that the system has enough generation resources to reach the anticipated LM lower limit. Constraint (a) ensures that the adjustments are only reductions of the originally set LVRT setting. Constraint (b) specifies that the LM of the system after the adjustments satisfies the aforementioned equilibrium/solution existence at $\lambda = \lambda_{\text{limit}}$.

It is to be noted that even after adjusting the LVRT settings, if an RG trips offline before reaching $\lambda_{\text{limit}}$, it is as good as not adjusting the LVRT settings in the first place. Therefore, the aim of the method should be to identify RGs that are made to stay online until $\lambda_{\text{limit}}$ is reached. The LVRT settings to achieve these blockings are computed thereafter.

## IV. METHODOLOGY FOR ADAPTIVE LVRT ADJUSTMENT

The problem defined in (3) is a nonlinear mix-integer programming problem. In order to solve it efficiently, we propose a method that successively screens out infeasible/ineffective candidate solutions in an attempt to focus on and evaluate only a handful of good solutions in greater detail. The method is composed of five steps (as shown in Figure. 2): 1) Run a full CPFLOW with RGs tripping to get the base P-V curve; 2) Find critical RGs in the system whose LVRT settings must be considered for adjustment; 3) Find effective combinations of RGs to be kept online/blocked from tripping offline to meet the LM limit; 4) Find LVRT adjustments necessary to achieve the previous step in order to rank feasible solutions; 5) Evaluate the effectiveness of top-ranked solutions using full CPFLOW and also the new LVRT settings needed. A detailed description of the main steps is presented next.

### A. Identifying Critical RGs

In the first stage, the aim is to find the group of RGs that tend to trip offline for the given load increase direction. The LVRT settings can be kept constant at all the other RGs, which reduces the starting number of candidate solutions. This is especially beneficial for applications to big systems where the search space of the solutions could be extremely large owing to the large number of RGs. For example, if there are 10 RGs installed in the system, the search space would be composed of $2^{10}$ candidate solutions. If only 5 of these 10 RGs are critical, namely, tending to trip, the search space could be decreased to $2^5$ candidate solutions which is much less than $2^{10}$. The voltage of each RG, $V_{est.}$, at $\lambda = \lambda_{\text{limit}}$ is estimated assuming a full quadratic relationship between $V$ and $\lambda$. RGs with $V_{est.} < c_{LVRT}$ are classified as critical.

### B. Screening for Feasible Solutions using Sensitivity

The RGs to be considered for LVRT adjustment consist of the identified critical RGs in the previous step and the RGs whose LVRT limits are violated in the base P-V curve. The LVRT settings of the rest of the RGs are left unchanged. Given the base P-V curve, we estimate the changes in LM due to blocking RGs from tripping. Since the RGs are modeled as negative active loads, we conservatively model not changing LVRT setting (or not blocking) as loss of active power injection. The power flow equations are linearized at the SNB point of the base P-V curve. This process is shown in Figure. 3.

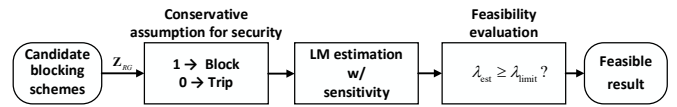

Figure. 3. Process of feasible solutions screening with sensitivity

Using the sensitivity-based method proposed in [15], one can find the sensitivity of LM w.r.t. change in connection status of each RG. This sensitivity is obtained based on the derivative of the system power flow function at the SNB,

$$\left.\frac{\partial f}{\partial x}\right|_* dx + \left.\frac{\partial f}{\partial \lambda}\right|_* d\lambda + \left.\frac{\partial f}{\partial Z_{RG}}\right|_* dZ_{RG} = 0 \quad (5)$$

where $\bullet|_*$ indicates the derivative evaluated at the SNB. By pre-multiplying one left zero eigenvector $v|_*$ of $\left.\frac{\partial f}{\partial x}\right|_*$, the first term in the LHS is eliminated and the desired sensitivity is found.

$$d\lambda = -\left(v \left.\frac{\partial f}{\partial \lambda}\right|_*\right)^{-1} \left(v \left.\frac{\partial f}{\partial Z_{RG}}\right|_*\right) dZ_{RG} = A \, dZ_{RG} \quad (6)$$

The estimated LM $\lambda_{j,est}^*$ for the $j^{th}$ candidate solution is,

$$\lambda_{j,est}^* = \lambda_{base}^* + A \, \Delta Z_{RG} \quad (7)$$

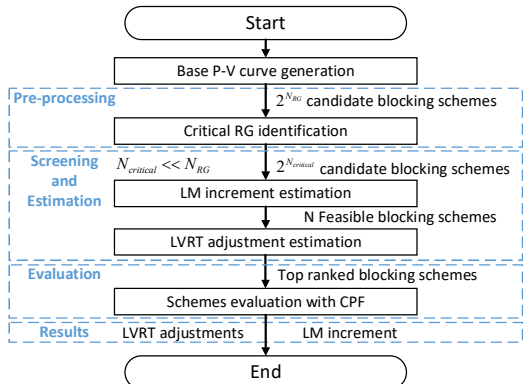

Figure. 2. Flow chart of the proposed method

where $\lambda_{base}^*$ refers to the LM of the base P-V curve; $\Delta Z_{RG} = Z_{RG,j} - Z_{RG,base}$ refers to the RG statuses' change from the final RG connecting statuses $Z_{RG,base}$ in the base case to the concerned candidate solution $Z_{RG,j}$. For each candidate solution, if $\lambda_{j,est}^* \geq \lambda_{\text{limit}}$, this candidate solution is considered feasible. It is to be noted that this sensitivity is evaluated at the SNB point of the base curve. Corresponding RGs are blocked from tripping when running the base case CPFLOW to prevent system collapse before reaching the true SNB.

### C. Estimating Adjustment of LVRT Settings

The aforementioned selection identifies the feasible RG blocking schemes. It is also necessary to know the amount of reduction in the LVRT settings that can realize the blocking in each feasible scheme. In order to make as little decrease as possible in the LVRT settings, to prevent sabotaging bulk system security, a method is proposed here to estimate the voltages of the PCCs of the RGs at $\lambda_{\text{limit}}$. These estimated voltages are used to find the reduction compared to the original LVRT settings. This iterative method starts from known equilibrium points on the base P-V curve which is the only accurate information one has. Candidate blocking schemes can also be mapped to the curves w.r.t. the RG connecting statuses. The input and output of this step are illustrated in Figure. 4.

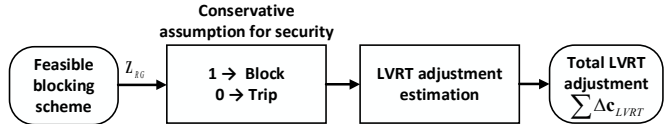

Figure. 4. Process of LVRT settings adjustment estimation

There are two scenarios to be considered depending on whether the base P-V curve has passed $\lambda_{\text{limit}}$ or not.

#### 1) The original curve has passed the LM limit

For each feasible solution, the Newton-Raphson method is used based on the power flow equations of the system with aimed RG connection statuses to be investigated, $Z_{RG,aimed}$. The starting point, $x_{base}$, of this iterative process is on the known base P-V curve, $Z_{RG,base}$. The calculation evolves along with state points on the P-V curves between the base curve and the aimed curve at the same load scale. Each of the curves can be mapped to a specific RG connection status combination. The result is the aimed state point, $x_{aimed}$, on the aimed P-V curve as shown in Figure. 5 (a).

#### 2) The original curve has not passed the LM limit

Under this circumstance, the starting state point, $x_{base}$, in scenario 1 does not exist as shown in Figure. 5 (b). Therefore, we take the SNB, $x_{nose,base}$, of the original curve to be the starting point. As this point is generally farther away from $x_{aimed}$, there is a larger chance that the Newton-Raphson based on the aimed connecting statuses encounters divergence. Hence, the varying process of the connecting statuses from the original ones to the aimed scheme is taken as a continuous process other than the previous discrete one. A middle point $x_{aimed,middle}$ on the aimed curve is introduced to be the goal of the iterative process. The reason is that this point has the largest chance to be both close to the aimed point, $x_{aimed}$, and on the aimed curve. The tangent estimation is utilized here based on the gradient on $x_{aimed,middle}$ to find the estimated $x_{aimed}$.

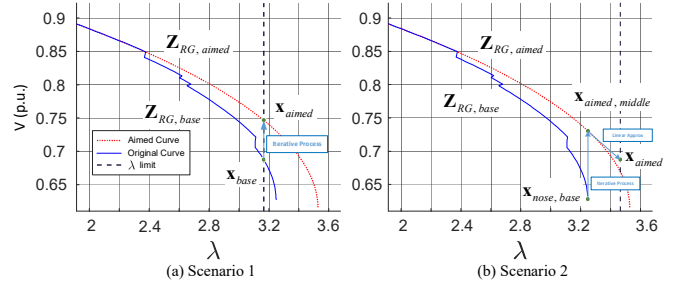

Figure. 5. RG voltage magnitude estimation on feasible curves

With the $V$ in $x_{aimed}$ estimated, the decrease of LVRT settings for the current blocking scheme is reached as follows,

$$\Delta c_{LVRT} = max\{(c_{LVRT}^{original} - V_{RG}), 0\} \quad (8)$$

where $V_{RG} \in x_{aimed}$ is the vector of RG voltage magnitudes.

### D. Optimum Adjustment Scheme Selection and Evaluation

Considering the least adjustments of LVRTs for bulk power system security, $m$ feasible candidate solutions that have the minimum absolute summations of the estimated LVRT settings adjustments are selected for CPFLOW computation. This gives us the actual LVRT adjustment needed to realize the proposed blocking schemes. The one with the minimum adjustment is chosen as the best solution. The number of feasible candidate solutions chosen is a user-defined variable that depends on how many solutions users decide to investigate. The process of this step is illustrated in Figure. 6.

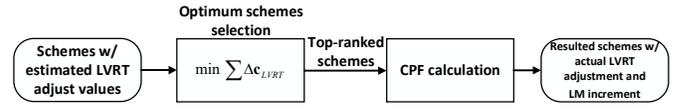

Figure. 6. Process of CPF evaluation

## V. CASE STUDY AND DISCUSSION

A modified IEEE 9-bus system with high RG penetration is used to demonstrate the effectiveness of the proposed method. There is one RG installed on each load bus (6 in total); their outputs and LVRT settings are given in Table I. Let the minimum desired LM be $\lambda_{\text{limit}} = 1.7$. In the base P-V curve, RG's LVRT limits are violated in the following order: bus $5 \rightarrow 9 \rightarrow 7 \rightarrow 4$. Notably, RG on bus 7 is blocked from tripping to reach SNB and thus to enable computing feasibility of candidate solutions in terms of LM increase as discussed in Section IV. B. The LM of the original system is 1.6723 after blocking the tripping of these two RGs, which is still less than $\lambda_{\text{limit}}$. Using the approach in Section IV. A, RGs on bus 5 and bus 9 are identified as critical. This can be confirmed in Figure. 7 where the predicted values are less than their respective LVRT limits. Since the LVRT limits on RGs at bus 4 and 7 are also violated in the base curve, according to Section IV. B, the complete group of RGs to be considered for adjustment are bus 4, 5, 7, and 9. The LVRT settings of RGs on bus 6 and 8 are kept constant due to the low likelihood of impacting LM.

The starting number of solutions is $2^6 - 1 = 63$. As four of the RGs are identified as critical, the actual number of candidate solutions is $2^4 - 1 = 15$, out of which 10 are deemed feasible based on the sensitivity-based screening logic developed in Section IV. B. Table II summarizes the various quantities estimated to select and rank top 5 ($m = 5$) feasible solutions. These are then studied in detail with CPFLOW to obtain the single best solution as well as the corresponding LVRT adjustments needed. The best solution (Solution 1, labeled red) requires actually lowering of the LVRT setting at bus 9 by 0.1178 to enhance the LM to 1.8134. This is greater than $\lambda_{\text{limit}}$ and thus feasible. The estimated results are close to the actual CPFLOW computation; this confirms the effectiveness of the proposed estimation methods. There does exist one solution that is found to be infeasible. This happens because of the usage of linear approximation of LM to estimate feasibility which is extremely fast but comes at a cost. However, as long as the best solution falls within the chosen five for detailed evaluation, it should not matter. There could also be coupling among the connection statuses of RGs. For example, one trips only if another trips offline. This can be seen in Solution 3 and 4 where Solution 3 aims to block bus 5 and 9, whereas Solution 4 keeps bus 4, 5, and 9 online. Results show bus 4 heavily depends on bus 5 and 9, and blocking bus 4 or not end up with same LVRT adjustment which is '0'. The best solution found through brute force CPFLOW calculations is the same as the first solution in Table II, which proves the effectiveness of the proposed method. The efficiency of the proposed approach is demonstrated from the fact that it needs to do CPFLOW computation six times (one for base P-V curve acquisition and five for evaluation of estimated solutions), whereas the brute-force method requires CPFLOW computation to be done 64 times.

TABLE I. RG SETTINGS OF STUDY CASE (P.U.)

| Bus Number | 4 | 5 | 6 | 7 | 8 | 9 |
|---|---|---|---|---|---|---|
| PG | 0.1 | 0.2 | 0.1 | 0.5 | 0.05 | 0.4 |
| LVRT const. | 0.8 | 0.9 | 0.8 | 0.83 | 0.8 | 0.8 |

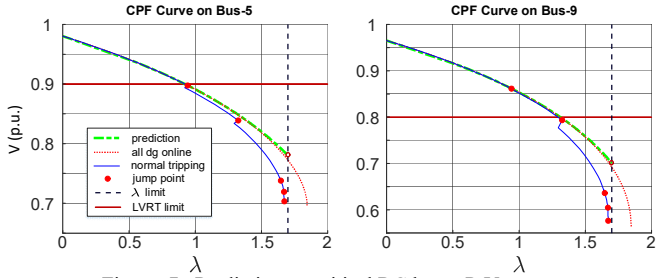

Figure. 7. Prediction on critical RG buses P-V curves

## VI. Conclusions

In this paper, we proposed a method to adaptively adjust the LVRT settings of the RGs as a preventive control mechanism to enhance voltage security. This novel methodology includes critical RG identification, feasible solution screening, LVRT adjustment estimation, and evaluation, and is developed to help the system attain the desired LM by preventing tripping of selected RGs by making minimum possible and only if absolutely necessary adjustment to their LVRT settings. The approach can be used both online to adapt the LVRT settings to changing system operating conditions and offline to figure out the areas requiring dynamic VAR support to minimize harmful RG tripping scenarios. The method is efficient in finding good quality solutions by avoiding the repetitive use of CPFLOW calculation, thereby enabling potential online applications. The proposed approach is deployed on a 9-bus test case. The study case results demonstrated the effectiveness and efficiency of the method. Future work will focus on deployment on large scale systems.

TABLE II. RG Blocking Solutions W/ Lowest Lvrt Adjustment

| Top Solution No. | | $\Delta c_{LVRT} = c_{LVRT}^{original} - c_{LVRT}^{new}$ | | | | LM |
|---|---|---|---|---|---|---|
| | | Bus 4 | Bus 5 | Bus 7 | Bus 9 | |
| 1 | Estimate | 0 | 0 | 0 | 0.1188 | 1.8198 |
| | Actual | 0 | 0 | 0 | 0.1178 | 1.8134 |
| 2 | Estimate | 0 | 0.1982 | 0.0299 | 0 | 1.7012 |
| | Actual | Infeasible Solution ($\lambda < \lambda_{\text{limit}}$) | | | | 1.6993 |
| 3 | Estimate | 0 | 0.1272 | 0 | 0.1072 | 1.8538 |
| | Actual | 0 | 0.1277 | 0 | 0.018 | 1.8450 |
| 4 | Estimate | 0 | 0.1286 | 0 | 0.109 | 1.8487 |
| | Actual | 0 | 0.1277 | 0 | 0.018 | 1.8450 |
| 5 | Estimate | 0.027 | 0.2007 | 0.0318 | 0 | 1.7063 |
| | Actual | 0.001 | 0.1746 | 0.0168 | 0 | 1.7033 |